# Smartphone-based point-of-care lipid blood test performance evaluation compared with a clinical diagnostic laboratory method

Kyongsik Yun[1], Juhee Lee[1], Jaekyu Choi[1], In-Uk Song[2], Yong-An Chung[3*]

[1]BBB Technologies Inc., 26, Samseong-ro 85-gil, Gangnam-gu, Seoul, South Korea
[2]Department of Neurology, Incheon St. Mary's Hospital, The Catholic University of Korea, 56 Dongsu-ro, Bupyeong-gu, Incheon, South Korea
[3]Department of Radiology, Incheon St. Mary's Hospital, The Catholic University of Korea, 56 Dongsu-ro, Bupyeong-gu, Incheon, South Korea
*Correspondence: yongan@catholic.ac.kr

**Abstract**
Managing blood lipid levels is important for the treatment and prevention of diabetes, cardiovascular disease, and obesity. An easy-to-use, portable lipid blood test will accelerate more frequent testing by patients and at-risk populations. We used smartphone systems that are already familiar to many people. Because smartphone systems can be carried around everywhere, blood can be measured easily and frequently. We compared the results of lipid tests with those of existing clinical diagnostic laboratory methods. We found that smartphone-based point-of-care lipid blood tests are as accurate as hospital-grade laboratory tests. Our system will be useful for those who need to manage blood lipid levels to motivate them to track and control their behavior.

## Introduction

Total cholesterol (TC), high density lipoprotein (HDL), and triglyceride (TG) levels indicate blood lipid levels. Controlling blood lipid levels is known to be associated with the treatment and prevention of various diseases, including diabetes mellitus (1), cardiovascular risk (2), and obesity (3). It is important to diagnose quickly and easily using a point-of-care-testing (POCT) device to effectively control blood lipid levels compared to hospital examinations (4).

By lowering the blood testing barriers, patients can check their blood more frequently. This gives you access to temporal changes in your biomarkers, making it easier to monitor and manage your health. One way to lower the test barriers is to use the familiar smartphone interface (5). We can take advantage of the sophisticated imaging technology and computing power available in smartphones. Therefore, the smartphone technology enables more accurate and comprehensive diagnostics. This will greatly improve preventive treatment for chronic metabolic diseases.

The elemark™ lipid check is a smartphone-based in-vitro diagnostic device for self-testing for rapid examination of three lipid markers. This device has a function to store / output the measured value. A whole blood sample that does not require centrifugation is used in the test to shorten the test time. This study examines the accuracy of the elemark™ lipid tests compared to hospital grade clinical diagnostic laboratory methods.

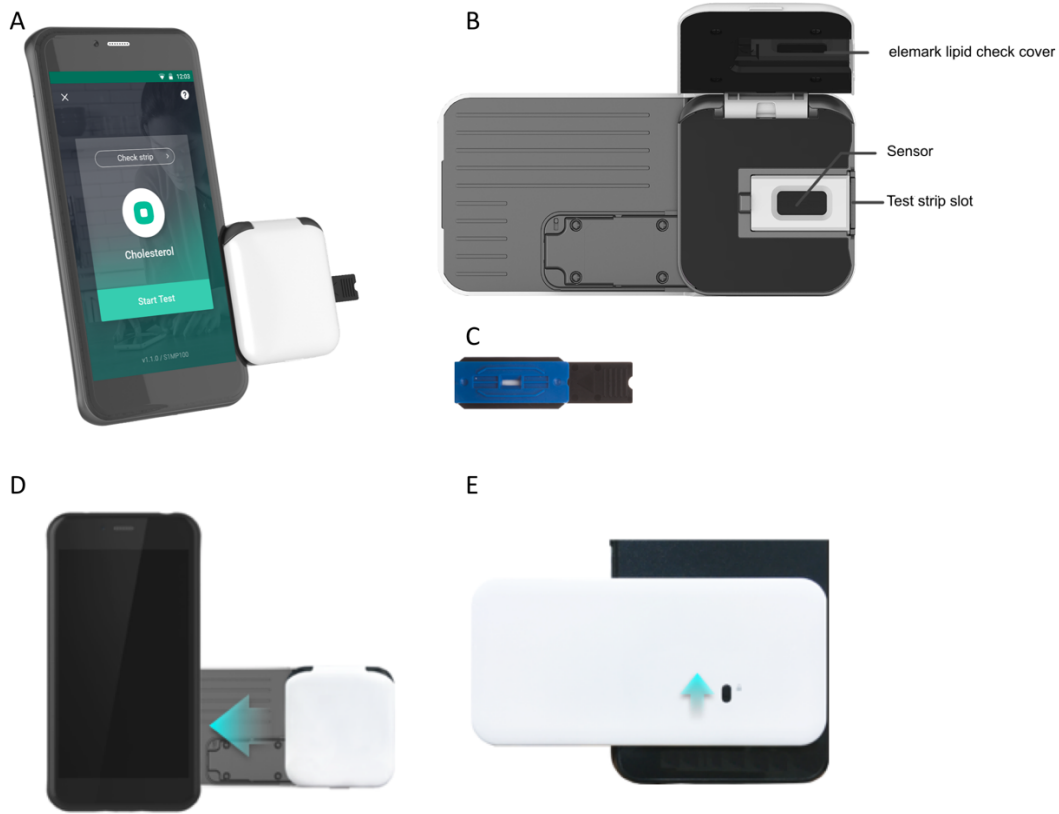

**Figure 1. Device components.** (A) elemark™, (B) the elemark™ lipid check, (C) the elemark™ lipid check cholesterol test strip, (D) connect the elemark™ lipid check to compatible mobile devices in the shown direction, (E) push the button up to lock the elemark™ lipid check.

**Methods**

The elemark™ lipid check device was developed in September 2016 as a self-testing cholesterol measuring device. TC, HDL and TG were measured using elemark™ compatible cholesterol test strips. The elemark™ System includes elemark™ Analyzer and SD LipidoCare™ lipid test strip (Figure 1).

This experiment was approved by the Incheon St. Mary's Hospital Institutional Review Board (IRB) in Incheon, South Korea, complied with the World Medical Association Declaration of Helsinki regarding ethical conduct of research involving human subjects. We followed the experimental procedures described in the normative references for the application of the International Organization for Standardization (ISO) (6,7). We anonymized the identity of the sample source. Because it is a simple blood test, the risk of the sample supplier is minimized. Minimal risk is defined as the degree and severity of harm or discomfort that may arise from studies not greater than the daily life of a healthy person or a routine physical or psychological examination (8).

The test temperature was controlled at 20 ~ 26 °C during the experiment. AU5800 Analyzer (Beckman Coulter Inc., IN, USA) was used as a reference device. Total cholesterol (TC), triglyceride (TG) and high density lipoprotein (HDL) were measured and reported. Low density lipoprotein (LDL) can be calculated as TC minus HDL minus TG / 5. LDL was not reported in this evaluation study.

Table 1. Sample distribution of total cholesterol (TC), triglyceride (TG), and high density lipoprotein (HDL) concentrations.

| TC concentration (mg/dL) | N | total | % |
|---|---|---|---|
| Below 200 mg/dL (Normal) | 86 | 116 | 74.1 |
| 200~239 mg/dL (Borderline High) | 24 | 116 | 20.7 |
| Above 240 mg/dL (High) | 6 | 116 | 5.2 |
| **TG concentration (mg/dL)** | **N** | **total** | **%** |
| Below 150 mg/dL (Normal) | 71 | 116 | 61.2 |
| 150~199 mg/dL (Borderline High) | 22 | 116 | 19.0 |
| 200~499 mg/dL (High) | 22 | 116 | 19.0 |
| Above 500 mg/dL (Very high) | 1 | 116 | 0.9 |
| **HDL concentration (mg/dL)** | **N** | **total** | **%** |
| Below 40 mg/dL (Low) | 22 | 116 | 19.0 |
| 40~59 mg/dL (Normal) | 60 | 116 | 51.7 |
| Above 60 mg/dL (High) | 34 | 116 | 29.3 |

The elemark™ cholesterol meter was stored at -30 °C to 70 °C, relative humidity of 90% or less and an altitude of less than 2000 meters. The cholesterol test strips (SD Biosensor LipidCare Lipid Profile Strips) were stored at room temperature between 2 °C and 32 °C and were used immediately after opening the individual pouches.

The venous whole blood in the treated ethylenediaminetetraacetic acid (EDTA) tube was used for elemark™, and the venous serum blood in the treated EDTA tube was separated from the venous whole blood and used in the reference device (AU5800). We recruited 116 participants for a blood test (809 women, 36 men, 70.8 ± 10.8 years) (Table 1). Samples were tested and analyzed within one day of blood collection. The comparison results were analyzed between elemark™ and AU5800. The correlation coefficient was used to evaluate the correlation between the two measurements (Figure 2). A detailed elemark™ user manual is attached as an attachment (supporting information).

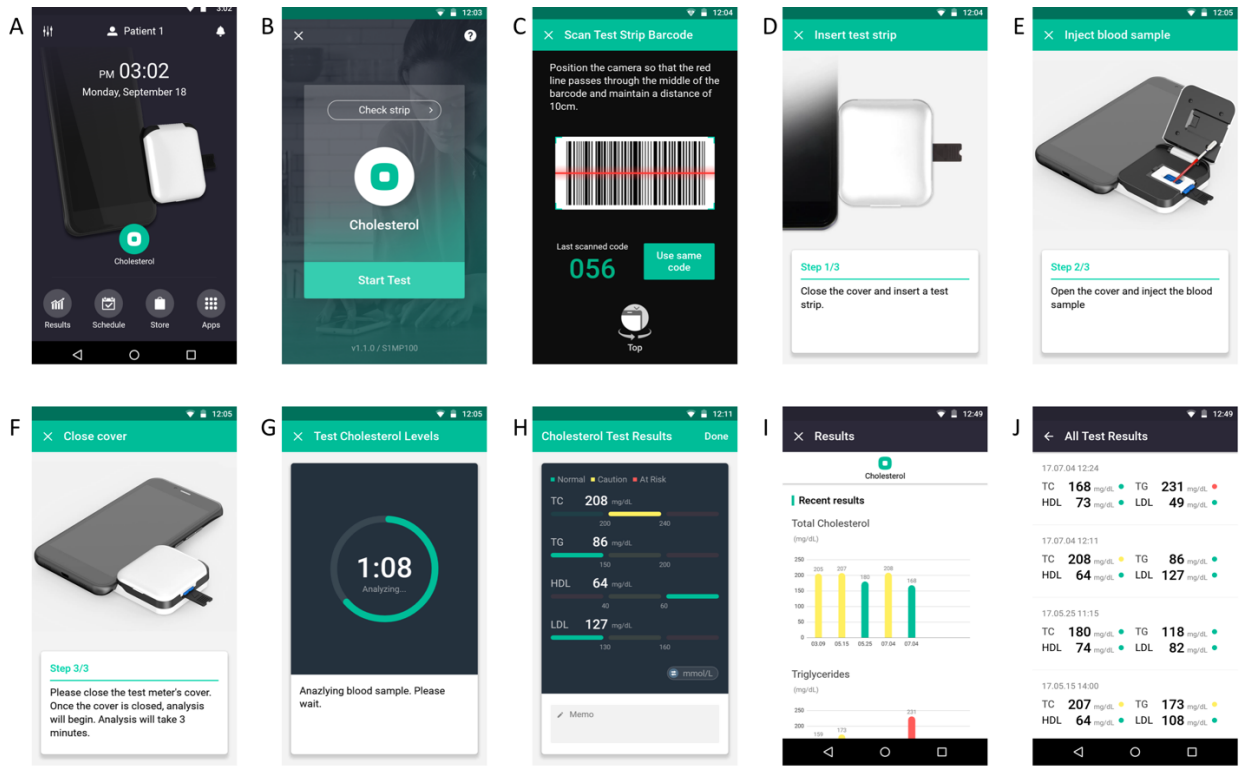

**Figure 2. Test procedure.** (A) Tap on the cholesterol icon to launch the app. (B) Press the Start Test button. (C) Use the camera of the mobile device to scan the barcode in the strip box. Once the cholesterol test strip is validated, the app automatically goes to the next step. (D) Insert the cholesterol measurement strip into the elemark™ lipid check as shown. (E) Blood samples are collected using a lancing device. Open the elemark™ lipid check cover, place the open end of the capillary into the sample area of the cholesterol test strip, and gently squeeze the capillary tube bulb to inject the blood sample. (F) Follow the on-screen instructions to close the cover. (G) Wait three minutes for analysis. (H) When the analysis is finished, the test results are displayed on the screen. (I) The user can check the latest test result graphically. (J) The user can check all test results.

Tukey's fences were used to identify outliers based on interquartile range. The interquartile range is a measure of the statistical variance and is equivalent to the difference between 75th and 25th percentile. For example, if *Q1* and *Q3* are the lower and upper quartiles, respectively, you can define anomalies with any observation outside the range:

$$[Q1 - k(Q3 - Q1), Q3 + k(Q3-Q1)] \quad (1)$$

When k = 1.5, values out of the above range were regarded as outliers (9,10).

**Results**

We compared the results between the elemark™ and reference devices (Table 2). Linear regression analysis showed that TC, TG, and HDL had a high correlation between the two devices (TC: correlation coefficient (R) = 0.97, coefficient of determination ($R^2$) = 0.94, p-value = 2.25X10$^{-73}$; TG: R = 0.99, $R^2$ = 0.98, p-value = 1.34X10$^{-92}$; HDL: R = 0.97, $R^2$ = 0.93, p-value = 1.67X10$^{-69}$).

Table 2. System accuracy

|  | <100mg/dL | | | ≧100mg/dL | | | |
| --- | --- | --- | --- | --- | --- | --- | --- |
|  | Within ±5mg/dL | Within ±10mg/dL | Within ±15mg/dL | Within ±5% | Within ±10% | Within ±15% | Within ±20% |
| TC | N/A (Device range: 100~450mg/dL) | | | 87/116 75.0% | 112/116 96.6% | 115/116 99.1% | 116/116 100% |
| TG | 8/24 33.3% | 17/24 70.8% | 23/24 95.8% | 42/92 45.7% | 71/92 77.2% | 89/92 96.7% | 92/92 100% |
| HDL | 98/116 84.5% | 116/116 100% | 116/116 100% | N/A (Device range: 25~95 mg/dL) | | | |

The system accuracy results showed that 99.1% of the TC concentration values at 100mg/dL or above were within ±15% of the reference results, 96.7% of the TG concentration values at 100mg/dL or above were within ±15% of the reference results, and 100% of the HDL concentration values less than 100mg/dL were within ±10mg/dL of the reference results (Table 2).

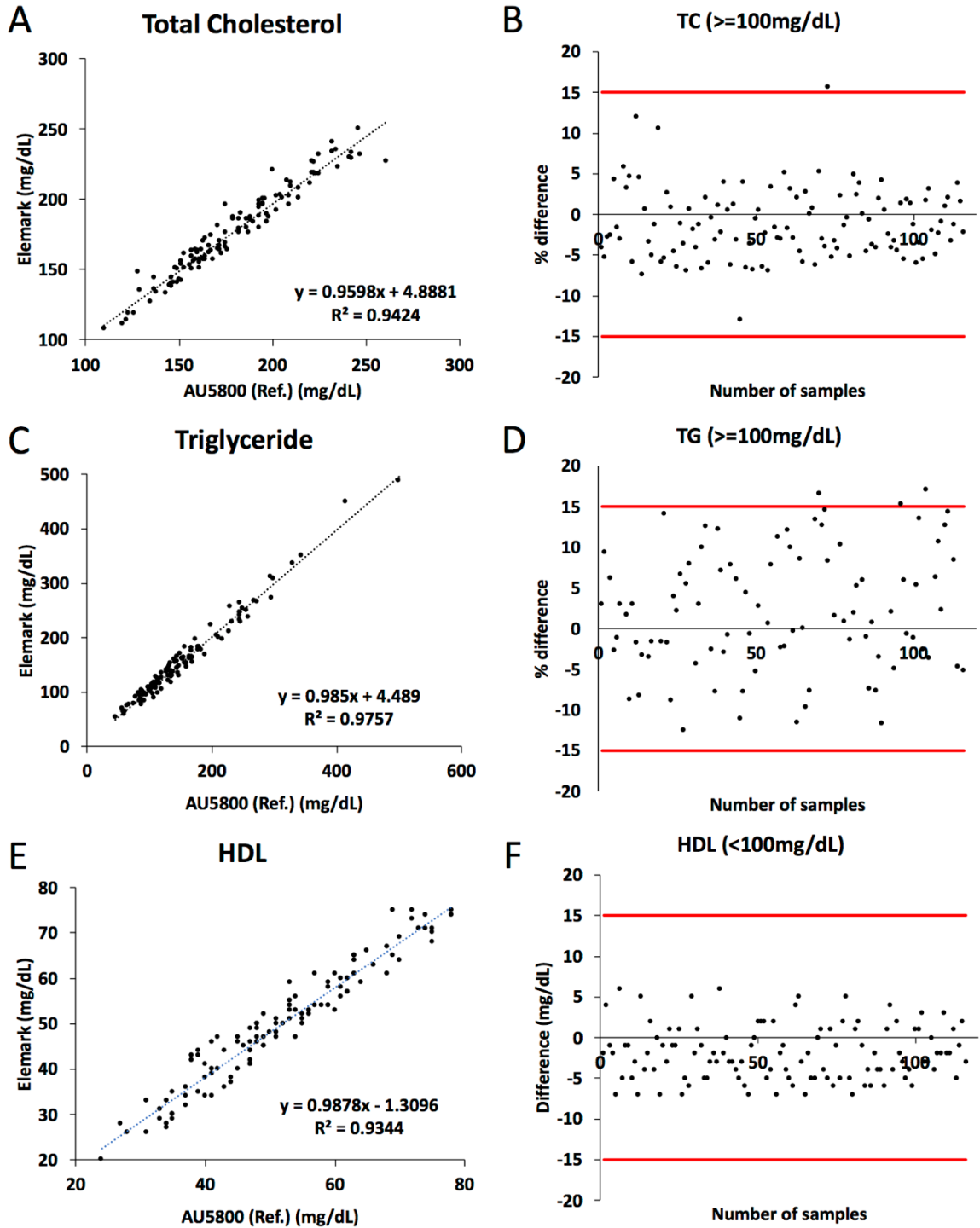

**Figure 3. Regression and difference charts.** (A) Total cholesterol (TC) regression chart, (B) percept difference chart with ±15% accuracy standard, (C) triglyceride (TG) regression chart, (D) percent difference chart with ±15% accuracy standard, (E) high density lipoprotein (HDL) regression chart, (F) percent difference chart with ±10% accuracy standard.

**Discussion**

We verified the accuracy of elemark™ by confirming the correlation between the whole blood lipid measurement values of three lipid markers (TC, TG, HDL) from elemark™ and the serum lipid values of the same three lipid markers from existing automated hospital equipment. elemark™ satisfied the acceptance criteria for the CE approval; 95% of the individual blood-testing parameter result shall fall within ±10mg/dL of the appropriate reference result at HDL concentration<100mg/dL and within ±15% of the reference result at TC and TG concentration ≥100mg/dL (11).

We used a smartphone-based easy-to-use and intuitive user interface (Figure 2) (12,13). Smartphone-based blood tests have several advantages in terms of computation, communication, and imaging. In addition, data generated from point-of-care devices can be easily shared with caregivers and healthcare professionals, helping to manage chronic disease in patients. Because the device is always connected to the network, data points cannot be lost, and the generated data can be automatically saved and configured for later analysis. Future machine learning-based analysis will allow us to predict cholesterol levels and calculate the risk of metabolic disease (14).

Our study has several limitations. We used venous blood for the elemark™ test. Capillary whole blood results are known to be different from venous whole blood results (15). Future studies showing actual use cases should use capillary whole blood. We should also be aware that the recruited participants were mainly from the elderly population (age: 70.8 ± 10.8). However, we believe that our current study of method comparison results is independent of age distribution. Rather, our results showed that elderly participants at high risk for cardiovascular disease produced accurate cholesterol diagnostics (16).

**Acknowledgment**

This study was supported by BBB Technologies Inc..